\documentstyle[a4,11pt]{article} 

\input amssym.def       
\input amssym.tex

\def\double{\Bbb}

\def\cc{{\double C}}     
\def\nn{{\double N}}       
\def\zz{{\double Z}}

\def\rr{{\double R}}

\newtheorem{theorem}{Theorem}
\newtheorem{lemma}[theorem]{Lemma}

\newtheorem{definition}[theorem]{Definition}
\newtheorem{proposition}[theorem]{Proposition}

\newtheorem{prodef}[theorem]{Proposition-Definition}

\newcommand{\be}{\begin{equation}}
\newcommand{\ee}{\end{equation}}
\newcommand{\beq}{\begin{eqnarray}}
\newcommand{\eeq}{\end{eqnarray}}
\newcommand{\om}{\omega}
\newcommand{\Om}{\Omega}

\def\nat{\natural}
\def\id{\mbox{Id}}

\newcommand{\la}{\lambda}

\newcommand{\non}{\nonumber}

\newcommand{\Tr}{\mbox{Tr}}

\newcommand{\te}{\theta}

\def\d{\partial}

\def\im{\mbox{Im}}
\def\ker{\mbox{Ker}}

\begin{document}

\begin{center}

{\large BRS-CHERN-SIMONS FORMS AND CYCLIC HOMOLOGY}
\vskip 1cm
{\bf Denis PERROT\footnotemark[1]}
\vskip 0.5cm
Centre de Physique Th\'eorique, CNRS-Luminy,\\ Case 907, 
F-13288 Marseille cedex 9, France \\[2mm]
{\tt perrot@cpt.univ-mrs.fr}
\end{center}
\vskip 0.5cm
\begin{abstract} 
We use some BRS techniques to construct Chern-Simons forms generalizing the 
Chern character of $K_1$ groups in the Cuntz-Quillen description of cyclic 
homology.
\end{abstract}

\vskip 0.5cm

\noindent {\bf MSC91:} 19D55, 81T13, 81T50\\


\footnotetext[1]{Allocataire de recherche MENRT.}

\noindent{\bf I. Introduction}\\

The index theorem for families has proven to be useful in the study of quantum 
anomalies and related topics \cite{F,Si}. In a previous paper we gave a simple 
formula expressing certain BRS cocycles as generalised Chern-Simons forms 
involving an (equivariant) family of Dirac operators \cite{P}. These forms are 
just de Rham cocycles on a parameter space. In the present paper we show that a 
similar construction of BRS-Chern-Simons forms yields cyclic homology classes 
generalizing the Chern character of $K_1$-groups.\\
This is achieved in the Cuntz-Quillen formalism for cyclic (co)homology, since 
in this framework Chern character forms are easily constructed from connections 
and curvatures. In order to avoid analytical technicalities with Dirac operators 
we focus essentially on the algebraic setting here.\\

The paper is organized as follows. After recalling the definition of 
$X$-complexes, we construct the Chern-Simons cycle in section III. Then we use 
the description of cyclic homology in terms of projective limits of 
$X$-complexes \cite{CQ} to obtain the result in the last section.

\vskip 1cm

\noindent{\bf II. $X$-complexes}\\

We first recall some basic notions \cite{CQ}. Let $R$ be a unital associative 
algebra over $\cc$. The space of universal one-forms is the tensor product 
\be
\Om^1(R)=R\otimes\overline{R}\qquad\mbox{with} \quad \overline{R}=R/\cc\ .
\ee
We adopt the notation $xdy$ for the element $x\otimes \overline{y}$, $x,y\in R$. 
In particular one has $d1=0$. $\Om^1(R)$ is naturally an $R$-bimodule, with left 
multiplication $x(ydz)=(xy)dz$, and the right multiplication comes from the 
Leibniz rule
\be
(dx)y := d(xy)-xdy\ .
\ee
The space of $n$-forms is obtained by tensoring over $R$:
\be
\Om^n(R)=\underbrace{\Om^1(R)\otimes_R\Om^1(R)...\otimes_R\Om^1(R)}_{n \ times}\ 
.
\ee
$\Om^n(R)$ is an $R$-bimodule isomorphic to 
$R\otimes\overline{R}\otimes...\otimes\overline{R}$ (tensor products over 
$\cc$). The elements of $\Om^n(R)$ will thus be denoted $x_0dx_1...dx_n$. This 
together with the space of zero-forms $\Om^0(R)=R$ implies that the direct sum 
\be
\Om(R)=\bigoplus_{n\ge 0} \Om^n(R)
\ee
is a unital differential graded (DG) algebra as follows. The multiplication 
$\Om^n\times \Om^m\rightarrow \Om^{n+m}$ is given by tensoring over $R$, and the 
differential $d:\Om^n\rightarrow\Om^{n+1}$ is simply
\be
d(x_0dx_1...dx_n)=dx_0dx_1...dx_n\ ,
\ee
with $d1=0$ for $1\in R=\Om^0(R)$. Clearly $d^2=0$ and the Leibniz rule holds:
\be
d(\om_1\om_2)=d(\om_1)\om_2 + (-)^n \om_1 d\om_2 \qquad \om_1 \in \Om^n(R)\ .
\ee

Next, the Hochschild operator $b: \Om^n(R)\rightarrow \Om^{n-1}(R)$,
\beq
b(\om dx)&=& (-)^{n-1}[\om,x]\non\\
b(x) &=&0 \qquad \mbox{for}\quad \om\in\Om^{n-1}(R)\ ,\ x\in R\ ,
\eeq
is well-defined because $[\om,1]=0$. It is easy to check that $b^2=0$. \\
Let us focus now on the $R$-bimodule $\Om^1(R)$. We let $\nat$ be the projection 
onto its abelianization 
\be
\Om^1(R)_{\nat}=\Om^1(R)/[R,\Om^1(R)]\ ,
\ee
so that the image of $xdy$ under the map $\nat: \Om^1(R)\rightarrow 
\Om^1(R)_{\nat}$ is written $\nat xdy$. Since $b^2=0$ the map
\beq
b:\Om^1(R)&\rightarrow& R \non\\
xdy &\mapsto& [x,y]
\eeq
vanishes on $\Om^1(R)\cap\im\ b = [R,\Om^1(R)]$, and thus defines an operator on 
the quotient
\beq
\overline{b}:\Om^1(R)_{\nat}&\rightarrow& R\non\\
\nat xdy &\mapsto& [x,y]\ .
\eeq
Consider also the operator
\beq
\nat d: R&\rightarrow& \Om^1(R)_{\nat}\non\\
x&\mapsto & \nat dx \ .
\eeq
Then $\nat d \circ \overline{b}=0$ and $\overline{b}\circ \nat d=0$, so that one 
gets a periodic complex
\be
...\to R \stackrel{\nat d}{\to}\Om^1(R)_{\nat}\stackrel{\overline{b}}{\to}R 
\stackrel{\nat d}{\to}\Om^1(R)_{\nat}\stackrel{\overline{b}}{\to}...\ .
\ee
\begin{definition}[\cite{CQ}] The $X$-complex of an algebra $R$ is the complex 
of period two
\be
X(R):\quad R \begin{array}{c}
\nat d \\[-1.5mm]
\longrightarrow\\[-3mm]
\longleftarrow\\[-1mm]
\overline{b}
\end{array} \Om^1(R)_{\nat}
\ee
with even degree $X_+(R)=R$ and odd degree $X_-(R)=\Om^1(R)_{\nat}$.
\end{definition}
\vskip 1cm

\noindent{\bf III. Chern-Simons forms}\\

We are ready to construct our Chern-Simons class in the odd homology 
$H_-(X(R))=\ker\ \overline{b}/\im\ \nat d$.\\

Let $L=L_+\oplus L_-$ be a $\zz_2$-graded unital algebra. The abelianization of 
$L$ is the quotient $L_{\nat}=L/[L,L]$ by graded commutators. The surjective map 
$\nat$ is the universal trace in the sense that any trace $\tau: L\to V$ with 
values in a vector space $V$ factorises through $L_{\nat}$, i.e. one has a 
commutative diagram
\be
\begin{array}{ccc}
 L   & \stackrel{\tau}{\to} & V \\
 & \searrow  &  \uparrow \\
 & & L_{\nat}
\end{array}
\ee
Tensoring the (bigraded) $X$-complex of $R$ by the bigraded space $L_{\nat}$ 
yields a new periodic complex
\be 
R \widehat{\otimes}L_{\nat} \begin{array}{c}
\nat d \otimes \id\\[-1mm]
\mbox{\LARGE $\longrightarrow$}\\[-2.5mm]
\mbox{\LARGE $\longleftarrow$}\\[-1mm]
\overline{b}\otimes\id
\end{array} \Om^1(R)_{\nat}\widehat{\otimes}L_{\nat}\ ,
\ee
where $\widehat{\otimes}$ is the graded tensor product. We shall construct a 
cycle in the odd part $\Om^1(R)_{\nat}\otimes (L_{\nat})_+$ starting from the 
following data:
\begin{itemize}
\item Two odd elements $D$ and $\te$ of $L$, which we call ``Dirac operator'' 
and ``connection'' respectively.
\item An invertible even element $g\in R\otimes L_+$.
\end{itemize}
We work in the graded tensor product $\Om(R)\widehat{\otimes}L$, which is a DG 
algebra for the differential $d\otimes \id$. By abuse of notation, the elements 
$1\otimes D$ and $1\otimes\te$ of $\Om(R)\widehat{\otimes}L$ are still written 
$D$ and $\te$.\\
Define the gauge transform of $\te$:
\be 
\te^g= g^{-1}\te g+g^{-1}[D,g]\qquad \in R\otimes L_-\ ,
\ee
and the Maurer-Cartan form
\be
\om= g^{-1}dg\qquad \in \Om^1(R)\otimes L_+\ .
\ee
They fulfill the usual BRS relations \cite{MSZ}
\be
d\te^g= -[D+\te^g, \om]\ ,\qquad d\om=-\om^2\ ,
\ee
since $dD=d\te=0$. Fix $t\in [0,1]$ and consider the superconnection \cite{Q1}
\be
\nabla= d+D+t(\om+\te^g)
\ee
with curvature
\be
\nabla^2= (t^2-t)(\om^2+[\om,\te^g]) +(D+t\te^g)^2 \qquad \in 
(\Om(R)\widehat{\otimes}L)_+
\ee
For $t=0,1$ the curvature satisfies the horizontality condition (or ``Russian 
formula'') $\nabla^2\in R\otimes L_+$:
\be
\nabla^2_{t=0}=D^2\ ,\qquad \nabla^2_{t=1}=(D+\te^g)^2\ .
\ee
Suppose now that the curvature is exponentiable in $\Om(R)\widehat{\otimes}L$, 
i.e. the series $e^{-\nabla^2}=\sum_n \frac{(-)^n}{n!}\nabla^{2n}$ makes sense. 
For this we have to consider suitable topological algebras, but in virtue of the 
algebraic construction presented here, we only have to assume that 
exponentiation exists and that the Duhamel formula holds:
\be
e^{(A+B)}= e^A +\sum_{n\ge 1} \int_{\Delta_n} e^{u_0A}Be^{u_1A}...Be^{u_nA}
\ee
where $\Delta_n$ is the $n$-simplex $\{(u_0,...,u_n)\in [0,1]^{n+1}|\sum_i 
u_i=1\}$. Put 
\be
\mu=(\om+\te^g)e^{-\nabla^2}\qquad\in (\Om(R)\widehat{\otimes}L)_-
\ee
and define $\mu_1$ as the component of $\mu$ in $\Om^1(R)\otimes L_+$:
\be
\mu_1= \om e^{-(D+t\te^g)^2} + (t-t^2)\te^g\int_0^1 du\, e^{-u(D+t\te^g)^2} 
[\om,\te^g] e^{(u-1)(D+t\te^g)^2}\ ,
\ee
where we used the Duhamel expansion with $A=-(D+t\te^g)^2$ and 
$B=(t-t^2)(\om^2+[\om,\te^g])$, retaining only the terms of degree one in $\om$. 
Then composing with the trace
\be
\nat\otimes\nat : \Om^1(R)\widehat{\otimes}L \to 
\Om^1(R)_{\nat}\widehat{\otimes}L_{\nat}
\ee
and integrating over $t\in [0,1]$ we obtain the Chern-Simons form:
\begin{definition}  The Chern-Simons form associated to the triple $(D,\te,g)$ 
is
\be
cs= \nat\otimes\nat \int_0^1 dt\, \mu_1\qquad \in 
\Om^1(R)_{\nat}\widehat{\otimes}L_{\nat}\ .
\ee
\end{definition}
Observe that if we put $D=\te=0$ then $cs$ is simply the Cartan form 
$\nat\otimes\nat\, \om$. It is closed with respect to $\overline{b}\otimes\id$, 
indeed
\be
(\overline{b}\otimes\id)(\nat\otimes\nat\, g^{-1}dg)=(\id\otimes\nat) 
([g^{-1},g])=(\id\otimes\nat)(1-1)=0\ .
\ee
The following proposition shows that in general $cs$ is a deformation of the 
Cartan cycle: 

\begin{proposition} i) $cs\in \Om^1(R)_{\nat}\widehat{\otimes}L_{\nat}$ is 
closed for $\overline{b}\otimes\id$.\\
ii) Its homology class in $H_-(X(R)\widehat{\otimes}L_{\nat})$ is equal to that 
of $\nat\otimes\nat\, \om$. In particular it does not depend on $D$ and $\te$.
\end{proposition}
{\it Proof:} We use a two-parameters transgression as in \cite{Q2} Thm 2. Let 
$t,u\in\rr$ and denote $\Om_{DR}(\rr^2)$ the de Rham complex corresponding to 
$t,u$. We work in the DG algebra 
$\Om_{DR}(\rr^2)\widehat{\otimes}\Om(R)\widehat{\otimes}L$ with total 
differential $d+dt\d_t+du\d_u$. Consider the superconnection
\be
\widetilde{\nabla}=d+dt\d_t+du\d_u +\rho \ ,\qquad \rho=uD+t(\om+u\te^g)\ ,
\ee
with curvature
\beq
{\widetilde{\nabla}}^2&=&d\rho +\rho^2 +dt\d_t\rho +du\d_u\rho \non\\
&=& \nabla^2 +dt\d_t\rho +du\d_u\rho \ .
\eeq
One has
\be
\nabla^2= t(d\om+ud\te^g)+(uD+t(\om+u\te^g))^2\ .
\ee
Since $d\te^g=-[D+\te^g,\om]$ the computation for $t=0,1$ gives
\be
\nabla^2_{t=0}=u^2D^2\ ,\qquad \nabla^2_{t=1}=u^2(D+\te^g)^2\ .
\ee
Now using the Duhamel formula we develop $e^{-{\widetilde{\nabla}}^2}$ in powers 
of $dt,du$:
\be
e^{-{\widetilde{\nabla}}^2}= e^{-\nabla^2} - dt\, \mu -du\,\nu +dt\,du\la\ ,
\ee
with
\be
\mu=\int_0^1ds\, e^{-s\nabla^2}\d_t\rho\, e^{(s-1)\nabla^2}\ ,\qquad
\nu=\int_0^1ds\, e^{-s\nabla^2}\d_u\rho\, e^{(s-1)\nabla^2}\ .
\ee
Define $\Om(R)_{\nat}=\Om(R)/[\Om(R),\Om(R)]$ and write $\nat$ for the trace
\be
\nat: \ \Om_{DR}(\rr^2)\widehat{\otimes}\Om(R)\widehat{\otimes}L\to 
\Om_{DR}(\rr^2)\widehat{\otimes}\Om(R)_{\nat}\widehat{\otimes}L_{\nat}\ .
\ee
Note that the total differential is still defined on the range of $\nat$. The 
Bianchi identity $[\widetilde{\nabla},{\widetilde{\nabla}}^2]=0$ implies the 
usual closure of the Chern character form
\be
(d+dt\d_t+du\d_u)\nat e^{-{\widetilde{\nabla}}^2}= \nat 
[\widetilde{\nabla},e^{-{\widetilde{\nabla}}^2}]=0\ .
\ee
Keeping the term proportional to $dt\,du$ in this expression gives
\be
\nat\d_u\mu -\nat \d_t\nu+\nat d\la=0 \ .\label{toto}
\ee
Explicitly we have
\be
\nat\mu = \nat ((\om+u\te^g)e^{-\nabla^2})\ ,\qquad \nat\nu= 
\nat((D+t\te^g)e^{-\nabla^2})\ .
\ee
Integrating eq. (\ref{toto}) over $dt$ yields
\beq
\d_u\int_0^1 dt\,\nat (\om+u\te^g)e^{-\nabla^2}&=& \nat\nu \Big|_{t=0}^{t=1} - 
\int_0^1 dt\, \nat d\la \\
&=& \nat (D+\te^g)e^{-u^2(D+\te^g)^2} -\nat De^{-u^2D^2} -\int_0^1 dt \nat d\la\ 
.\non
\eeq
Let $p_n$ be the projection of $\Om(R)_{\nat}\widehat{\otimes}L_{\nat}$ onto 
$\Om^n(R)_{\nat}\widehat{\otimes}L_{\nat}$ and set
\be
cs(u):=p_1 \int_0^1 dt\,\nat (\om+u\te^g)e^{-\nabla^2}\ ,\qquad \la_0:=p_0\la\ .
\ee
One has
\be
\d_ucs(u)= -\int_0^1 dt\, \nat d\la_0\ .
\ee
But $cs(0)=\nat\om$ and $cs(1)=cs$ so that 
\be
cs=\nat \om - \nat d\int_0^1du\int_0^1 dt\, \la_0\ ,
\ee
and the conclusion follows. $\Box$\\

The proposition thus implies that the form
\be
cs= \nat\otimes\nat \int_0^1 dt\, \mu_1
\ee
is $\overline{b}$-closed. For the applications we have in mind, $L$ is an 
algebra of operators in a bigraded (infinite-dimensional) Hilbert space, and $D$ 
is an ordinary Dirac operator. Then $\mu_1$ involves the heat kernel $e^{-D^2}$, 
which we use as a regulator for the ordinary supertrace $\mbox{Tr}_s$ with 
values in $\cc$. Then we can compose the universal trace $\nat: L\to L_{\nat}$ 
by $\mbox{Tr}_s$ to get a cycle
\be
\nat\otimes\mbox{Tr}_s \int_0^1 dt (\om e^{-(D+t\te^g)^2} + (t-t^2)\te^g\int_0^1 
du\, e^{-u(D+t\te^g)^2} [\om,\te^g] e^{(u-1)(D+t\te^g)^2})
\ee
in the odd part of the $X(R)$. This is an analogue of the BRS cocycles obtained 
in \cite{P} by transgression of the Chern character of families of Dirac 
operators. Since we don't want to treat the analytic details here we will focus 
only on the purely algebraic (and universal) side.

\vskip 1cm
\noindent{\bf IV. Cyclic homology}\\

In this section we use freely some results of Cuntz and Quillen concerning the 
$X$-complex description of cyclic homology. For convenience the material we need 
is briefly reviewed, but we strongly refer to \cite{CQ} for details.\\

Let $R$ be an associative unital algebra, $I$ an ideal of $R$. For any $n\ge 1$, 
$R/I^n$ is a unital algebra, thus we get a sequence of surjective maps
\be
...\to R/I^{n+1}\to R/I^n\to ...\to R/I\ .
\ee
By definition the $I$-adic completion of $R$ is the corresponding projective 
limit
\be
\widehat{R}= \lim_{\longleftarrow} R/I^n \ .
\ee
Passing to the $X$-complexes, the projection $R/I^{n+1}\to R/I^n$ induces a 
surjective map of complexes $X(R/I^{n+1})\to X(R/I^n)$, whence a sequence
\be
...\to X(R/I^{n+1})\to X(R/I^n)\to ...\to X(R/I)\ ,
\ee
with projective limit
\be
\widehat{X}(R,I):= \lim_{\longleftarrow} X(R/I^n)\ .
\ee
The projections $\widehat{R}\to R/I^n$ for all $n$ induce maps of complexes 
$X(\widehat{R})\to X(R/I^n)$ which are compatible to each other in the sense 
that the diagram
\be
\begin{array}{ccc}
X(\widehat{R}) & \to & X(R/I^{n+1}) \\
 & \searrow  &  \downarrow \\
 & & X(R/I^n)
\end{array}
\ee
commutes. Thus one gets a canonical map of complexes
\be
X(\widehat{R})\to \widehat{X}(R,I)\ . \label{map}
\ee
The completion $\widehat{X}(R,I)$ is important in relation with the following 
fundamental example. Let $A$ be a unital algebra. Put $RA=\Om^+(A)$, the space 
of differential forms of even degree over $A$, endowed with the Fedosov 
(associative) product
\be
\om_1 \circ \om_2 = \om_1\om_2 - d\om_1 d\om_2\ ,\qquad \om_i\in\Om^+(A)\ .
\ee
The projection $RA\to A$ which sends a form to its component of degree zero, is 
an algebra homomorphism. By definition the ideal $IA$ is the kernel of this map. 
Explicitly,
\be
IA= \bigoplus_{n >0} \Om^{2n}(A)\ .
\ee
Then the $IA$-adic completion of $RA$ corresponds to the direct {\it product}
\be
\widehat{R}A = \prod_{n\ge 0} \Om^{2n}(A)
\ee
endowed with the Fedosov product $\circ$. The following fundamental result is 
due to Cuntz and Quillen:
\begin{theorem}[\cite{CQ}]
The even (resp. odd) homology of the complex $\widehat{X}(RA,IA)$ is isomorphic 
to the even (resp. odd) periodic cyclic homology of $A$:
\be
H_*(\widehat{X}(RA,IA))= HP_*(A)\ .
\ee
\end{theorem}
We are thus in a position to construct cyclic homology classes of $A$ by means 
of the universal Chern-Simons construction of the preceding section. Thus we 
consider the following data:
\begin{itemize}
\item A unital algebra $A$.
\item A $\zz_2$-graded unital algebra $L=L_+\oplus L_-$.
\item Two odd elements $D,\te\in L_-$.
\item An invertible even element $g\in A\otimes L_+$.
\end{itemize}
For $n\ge 0$ set $R_n=RA/IA^{n+1}$. As a space it is isomorphic to 
$\oplus_{k=0}^n \Om^{2k}(A)$. Thus $g\in A\otimes L_+$ can be considered as an 
element $g_n$ of $R_n\otimes L_+$.
\begin{lemma}
$g_n$ is invertible in the algebra $R_n\otimes L_+$, with inverse given by
\be
(g_n)^{-1}= \sum_{k=0}^n g^{-1}(dgdg^{-1})^k\qquad \mbox{in}\quad 
\bigoplus_{k=0}^n \Om^{2k}(A)\otimes L_+ \simeq R_n\otimes L_+\ ,
\ee
where $g^{-1}$ is the inverse of $g$ in $A\otimes L_+$.
\end{lemma}
{\it Proof:} See \cite{CQ} $\S$ 12. $\Box$\\

Thus for each $n$ the Chern-Simons form corresponding to the triple 
$(D,\te,g_n)$ gives an odd cycle in the complex $X(R_n)\widehat{\otimes} 
L_{\nat}$:
\be
cs_n \in H_-(X(R_n)\widehat{\otimes} L_{\nat})\ .
\ee
Moreover these cycles are compatible with the sequence of complexes
\be
...\to X(R_{n+1})\widehat{\otimes} L_{\nat}\to X(R_n)\widehat{\otimes} 
L_{\nat}\to ...\to X(R_0)\widehat{\otimes} L_{\nat}
\ee
in the sense that $cs_{n+1}$ maps to $cs_n$. One is led to the following
\begin{prodef}
The Chern character of the triple $(D,\te,g)$ is the homology class of the 
projective limit
\be
cs= \lim_{\longleftarrow} cs_n \quad \in H_-(\lim_{\longleftarrow} 
X(R_n)\widehat{\otimes} L_{\nat})\ .
\ee
Moreover this class does not depend on $D$ and $\te$.
\hfill $\Box$
\end{prodef}

The terminology ``Chern character'' is motivated by the following particular 
case. $L_+$ is the matrix algebra $M_N(\cc)$, for $N\in\nn$, and $L_-=0$. The 
subspace of commutators $[L,L]$ is equal to the set of traceless matrices, so 
that $L_{\nat}\simeq \cc$ and the universal trace $\nat$ corresponds to the 
ordinary trace $\Tr: M_N(\cc)\to \cc$. Thus $cs$ is an element of 
\be
H_-(\lim_{\longleftarrow} X(R_n))\simeq HP_1(A)\ ,
\ee
the odd periodic cyclic homology of $A$. Now $g$ is an invertible element of 
$M_N(A)$ and defines a class $[g]$ in the group $K_1(A)$. Introduce the limits
\beq
\widehat{g}&=&\lim_{\longleftarrow} g_n \qquad \in M_N(\widehat{R}A)\non\\
{\widehat{g}}^{-1}&=& \lim_{\longleftarrow} (g_n)^{-1}\ .
\eeq
For $D=\te=0$ one has simply
\be
cs_n = \nat\Tr\, ((g_n)^{-1}dg_n)\ ,
\ee
so that $cs$ is the image of
\be
\nat\Tr\,  ({\widehat{g}}^{-1}d\widehat{g})\quad \in H_-(X(\widehat{R}A))
\ee
under the map (\ref{map}). This corresponds precisely to the Chern character of 
$[g]$ in $HP_1(A)$ (cf. \cite{CQ}).

\vskip 1cm

\noindent{\bf Acknowledgments:} As usual, the author wishes to thank his advisor 
S. Lazzarini for his constant attention.

\makeatletter
\def\@biblabel#1{#1.\hfill}

\end{document}